\documentclass[aps,pre,onecolumn,superscriptaddress,nofootinbib]{revtex4-2}

\usepackage{amsmath,amssymb,amsfonts,mathtools}
\usepackage{graphicx}
\usepackage{bm}
\usepackage{physics}
\usepackage{hyperref}
\usepackage{xcolor}
\usepackage{subcaption}

\newcommand{\e}{\mathrm{e}}
\newcommand{\ii}{\mathrm{i}}
\newcommand{\RR}{\mathbb{R}}

\newcommand{\Lap}{\mathbf{L}}
\newcommand{\Adj}{\mathbf{A}}
\newcommand{\Deg}{\mathbf{D}}

\begin{document}

\title{Operational tracking loss in nonautonomous second-order oscillator networks}

\author{Ver\'onica Sanz}
\affiliation{Theoretical Physics Department, Universitat de Val\`encia, and Instituto de F\'isica Corpuscular (IFIC), CSIC--Universitat de Val\`encia, Spain}

\date{\today}

\begin{abstract}
We study when a network of coupled oscillators with inertia ceases to follow a time-dependent driving protocol coherently, using a simplified graph-based model motivated by inverter-dominated energy systems. We show that this loss of tracking is diagnosed most clearly in the frequency dynamics, rather than in phase-based observables. Concretely, a tracking ratio built from the frequency-disagreement observable \(E_\omega(t)\) and normalized by the instantaneous second-order modal decay rate yields a robust protocol-dependent freeze-out time whose relative dispersion decreases with system size. Graph topology matters substantially: the resulting freeze-out time is only partly captured by the algebraic connectivity \(\lambda_2\), while additional structural descriptors, particularly Fiedler-mode localization and low-spectrum structure, improve the explanation of graph-to-graph variation. By contrast, phase-sector observables develop strong non-monotonic and underdamped structure, so simple diagonal low-mode relaxation closures are not quantitatively reliable in the same regime. These results identify the frequency sector as the natural operational sector for nonautonomous tracking loss in second-order oscillator networks and clarify both the usefulness and the limits of reduced spectral descriptions in this setting.
\end{abstract}

\maketitle

\section{Introduction}
\label{sec:intro}

Synchronization in oscillator networks is often discussed in terms of the existence or stability of collective states under fixed conditions. In many systems of practical interest, however, the more relevant question is different: can the network \emph{track} a coherent state while couplings, injections, or operating conditions evolve in time? This question is especially natural in modern energy systems, where inverter-rich operation, reduced effective inertia, and reconfiguration protocols make nonautonomous collective dynamics increasingly relevant.

The purpose of this paper is to develop a reduced dynamical description of such tracking loss in a minimal setting. We do not attempt an engineering-grade model of a realistic grid. Instead, we ask which observables provide the cleanest and most robust operational signal of tracking loss in nonautonomous second-order oscillator networks, and to what extent those observables admit an interpretable reduced description.

A substantial literature has established the connection between oscillator-network synchronization and reduced power-grid models, and has clarified both synchronization and transient-stability questions in first- and second-order Kuramoto-type settings. In particular, D\"orfler and Bullo showed how reduced power-network dynamics can be related to coupled-oscillator models and derived influential synchronization criteria for complex oscillator networks and smart grids~\cite{dorfler_bullo_2011,dorfler_bullo_2013}. More broadly, the inertial or second-order Kuramoto model has long been used to study synchronization in settings where inertia and damping play an essential role, including power-network applications, and is known to display richer and qualitatively different behavior from the first-order model~\cite{filatrella_2008,dorfler_siam_2013,odor_2022,ha_2021_partial}. Recent work on renewable-dominated and converter-rich grids continues to stress the importance of frequency dynamics, reduced inertia, and multiscale nonlinear stability questions in such models~\cite{ma_2023_review,hartmann_2025}.

The contribution of the present paper is different in emphasis. We do not focus on synchronization thresholds, basin stability, or static operating conditions. Instead, we study \emph{nonautonomous coherent tracking} under time-dependent coupling protocols and ask which observables provide the cleanest operational indicator of tracking loss in second-order oscillator networks. Our main finding is that the relevant reduced observable is not primarily phase coherence but frequency disagreement: a tracking diagnostic built from \(E_\omega(t)\) and normalized by the instantaneous second-order modal decay rate remains informative precisely where phase-sector reductions become much less reliable.

Our starting point is a damped second-order phase dynamics on graphs with an explicitly time-dependent coupling protocol. This level is sufficient to incorporate inertia, damping, network structure, and nonautonomous driving while remaining tractable enough for modal analysis. The conceptual novelty of the paper is not the second-order network model by itself, but the construction of an operational frequency-sector diagnostic of nonautonomous tracking loss. More precisely, we show that the frequency-disagreement observable, when normalized by the instantaneous second-order modal decay rate, remains informative precisely in the regime where phase-sector reductions become much less reliable. Phase-sector observables such as phase disagreement and residual incoherence develop pronounced non-monotonic and underdamped behavior, so the simplest low-mode envelope ideas inherited from first-order dynamics are not quantitatively reliable there. By contrast, a diagnostic based on the frequency-disagreement observable \(E_\omega(t)\), normalized by the instantaneous second-order modal decay rate, gives a robust and protocol-sensitive crossover time.

The main results are the following. First, the frequency-based freeze-out time extracted from the normalized tracking ratio increases monotonically with the protocol timescale \(\tau\), with small run-to-run uncertainty. By freeze-out time we mean the operational crossover time at which the network ceases to track the changing coupling protocol coherently. Second, the relative dispersion of this operational freeze-out time decreases as the number of nodes grows, indicating that the diagnostic becomes sharper in larger networks. Third, graph topology matters: much of the graph-to-graph variation is organized by low-spectrum information, but the dependence is not exhausted by the algebraic connectivity \(\lambda_2\) alone. At intermediate ramps, additional structural descriptors such as Fiedler-mode localization improve the explanation substantially, while in the slower-ramp regime low-spectrum spacing becomes more relevant. Finally, although the real part of the instantaneous second-order modal exponent provides the correct normalization scale for the tracking diagnostic, the corresponding diagonal low-mode relaxation envelope is only quantitatively useful in the slow-ramp regime.

The paper is organized as follows. Section~\ref{sec:model} introduces the model, observables, and operational tracking diagnostic. Section~\ref{sec:modal} derives the second-order modal rate used throughout the paper. Section~\ref{sec:results_protocol} presents the protocol dependence and observable comparison. Section~\ref{sec:results_scaling} studies scaling with system size and graph structure, including the role of descriptors beyond \(\lambda_2\). Section~\ref{sec:envelopes} discusses the limits of diagonal low-mode relaxation envelopes. Section~\ref{sec:discussion} summarizes the interpretation and implications for future reduced models of inverter-motivated network dynamics.

\section{Model, observables, and operational tracking loss}
\label{sec:model}

\subsection{Nonautonomous second-order oscillator network}

We consider a network of \(N\) phase oscillators on an undirected graph with adjacency matrix \(\Adj\), with phases \(\theta_i(t)\in\RR\) and angular velocities \(\omega_i(t)=\dot\theta_i(t)\). The dynamics is
\begin{align}
\dot\theta_i &= \omega_i, \\
M\dot\omega_i &= P_i - D\omega_i - \sum_{j=1}^N K_{ij}(t)\sin(\theta_i-\theta_j),
\label{eq:model}
\end{align}
with uniform effective inertia \(M>0\) and damping \(D>0\). The static mismatch terms satisfy \(\sum_i P_i=0\). This should be regarded as a reduced oscillator model motivated by inverter-dominated and microgrid-like network dynamics, not as a full engineering model.

We focus on the simplest nonautonomous protocol,
\begin{equation}
K_{ij}(t)=K(t)A_{ij},
\end{equation}
with
\begin{equation}
K(t)=K_0\left(1+\frac{t}{\tau}\right)^{-\alpha},
\qquad K_0>0,\quad \tau>0,\quad \alpha>0.
\label{eq:Kprotocol}
\end{equation}
Here \(\tau\) controls the protocol timescale and \(\alpha\) its sharpness. The regime that proved most informative in the numerical analysis below is \(\alpha=3\), for which the changing coupling competes strongly with the intrinsic second-order transient.

\subsection{Graph structure and spectral quantities}

The graph Laplacian is
\begin{equation}
\Lap = \Deg - \Adj,
\end{equation}
with eigenvalues
\begin{equation}
0=\lambda_1 < \lambda_2 \le \lambda_3 \le \cdots \le \lambda_N.
\end{equation}
The algebraic connectivity \(\lambda_2\) sets the slowest linear spectral scale and remains an important organizing quantity throughout the paper. However, one of our conclusions is that the operational tracking-loss time is not fully reducible to \(\lambda_2\) alone in the second-order nonautonomous problem.

We study Erd\H{o}s--R\'enyi (ER), Watts--Strogatz (WS), and ring graphs. This allows us to contrast generic non-spatial graphs with a spatially constrained low-connectivity family that behaves anomalously.

\subsection{Collective observables}

The complex order parameter is
\begin{equation}
Z(t)=\frac{1}{N}\sum_{j=1}^N \e^{\ii\theta_j(t)} = R(t)\e^{\ii\Psi(t)},
\end{equation}
where \(R(t)\in[0,1]\) measures phase coherence and \(\Psi(t)\) is the collective phase.

Our phase-sector observables are the circular phase-disagreement energy
\begin{equation}
E_\theta(t)=\sum_{i=1}^N \mathrm{wrap}_{[-\pi,\pi)}\!\bigl(\theta_i(t)-\Psi(t)\bigr)^2,
\end{equation}
and the residual incoherence \(1-R(t)\).

The central observable of the present paper is the frequency-disagreement energy
\begin{equation}
E_\omega(t)=\sum_{i=1}^N \bigl(\omega_i(t)-\bar\omega(t)\bigr)^2,
\qquad \bar\omega(t)=\frac{1}{N}\sum_i \omega_i(t).
\label{eq:Eomega}
\end{equation}
The numerical results show that this frequency-sector quantity provides a much cleaner operational signal of tracking loss than the phase-sector observables in the second-order regime considered here.

\subsection{Operational tracking ratio}

The correct reference scale for the second-order problem is not the first-order rate \(K(t)\lambda_2\), but the real part of the slowest instantaneous modal exponent. For a linearized Laplacian mode \(c_\alpha\),
\begin{equation}
M\ddot c_\alpha + D\dot c_\alpha + K(t)\lambda_\alpha c_\alpha = 0,
\label{eq:modeeq}
\end{equation}
the instantaneous exponents are
\begin{equation}
\mu_{\alpha,\pm}(t)=\frac{-D \pm \sqrt{D^2-4MK(t)\lambda_\alpha}}{2M}.
\end{equation}
We define the reference decay rate as the real part of the slowest mode,
\begin{equation}
\Gamma_\alpha(t)=
\begin{cases}
\dfrac{D-\sqrt{D^2-4MK(t)\lambda_\alpha}}{2M}, & D^2\ge 4MK(t)\lambda_\alpha,\\[1.2ex]
\dfrac{D}{2M}, & D^2<4MK(t)\lambda_\alpha.
\end{cases}
\label{eq:Gammaalpha}
\end{equation}
This quantity is the natural second-order analog of the first-order relaxation rate and will be used throughout as a \emph{local} normalization scale.

Using the frequency-disagreement observable, we define the tracking ratio
\begin{equation}
r_\omega(t)= -\frac{1}{2}\frac{\partial_t \ln E_\omega(t)}{\Gamma_2(t)},
\label{eq:romega}
\end{equation}
where \(\Gamma_2(t)\) denotes \eqref{eq:Gammaalpha} evaluated at \(\lambda_2\). Numerically, \(\partial_t\ln E_\omega\) is estimated from a smoothed time series. In the fast-ramp case, the very early-time spike of \(r_\omega(t)\) should be interpreted as a launch-transient effect of the logarithmic derivative rather than as the physical freeze-out event itself; the extraction rule below is designed to identify the first clear post-peak loss of tracking.

An operational freeze-out time \(t_*\) is extracted from a first-post-peak rule: after a short initial transient, we identify the first peak of \(r_\omega(t)\) above a high threshold and define \(t_*\) as the first later time at which the ratio drops below a lower threshold for a sustained interval. This definition proved robust across repeated runs.

\section{Second-order modal scale and its limits}
\label{sec:modal}

The rate \(\Gamma_\alpha(t)\) provides a physically meaningful second-order modal scale and is essential for normalizing the tracking diagnostic. This point is strongly supported by the numerics: using the first-order scale \(K(t)\lambda_2\) leads to pathological behavior in the tracking ratio, whereas \eqref{eq:Gammaalpha} restores a sensible and protocol-sensitive operational crossover.

A stronger ansatz would be to use the corresponding integrated envelope,
\begin{equation}
\exp\!\left[-2\int_0^t \Gamma_\alpha(s)\,ds\right],
\label{eq:envelope}
\end{equation}
as a reduced relaxation model. Our numerical results show that this idea is only partly successful. In the frequency sector it becomes quantitatively reasonable in the slower-ramp regime, but it does not provide a uniformly accurate diagonal low-mode description across the full protocol range. In the phase sector, where \(E_\theta\) and \(1-R\) become strongly non-monotonic and underdamped, the same diagonal-envelope idea fails much more decisively.

Accordingly, the role of \(\Gamma_\alpha(t)\) in this paper is twofold. First, it provides the correct local second-order decay scale and therefore the proper normalization for the operational tracking ratio. Second, it offers an asymptotic guide to slow-ramp relaxation, but not yet a complete quantitative closure for all observables and protocols.

\section{Protocol dependence and observable selection}
\label{sec:results_protocol}

\begin{figure}[t]
    \centering
    \includegraphics[width=0.72\textwidth]{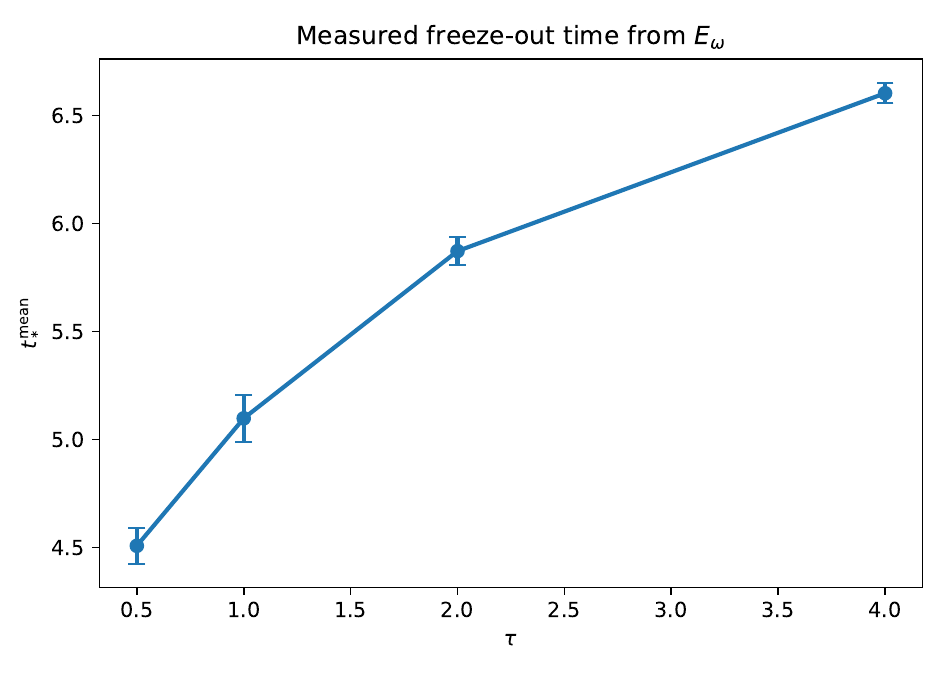}
    \caption{Measured freeze-out time extracted from the frequency-disagreement observable \(E_\omega\) as a function of the protocol timescale \(\tau\), using the second-order normalized tracking ratio \eqref{eq:romega}. Error bars show the standard error over realizations. The monotonic increase with \(\tau\) demonstrates that the frequency-based diagnostic is sensitive to the rate of nonautonomous driving.}
    \label{fig:tstar_protocol}
\end{figure}

We begin by asking whether the frequency-based diagnostic produces a clean and protocol-sensitive crossover time. Figure~\ref{fig:tstar_protocol} summarizes the resulting operational freeze-out time and shows a clear monotonic increase with the protocol timescale \(\tau\).

\begin{figure}[t]
    \centering
    \includegraphics[width=0.72\textwidth]{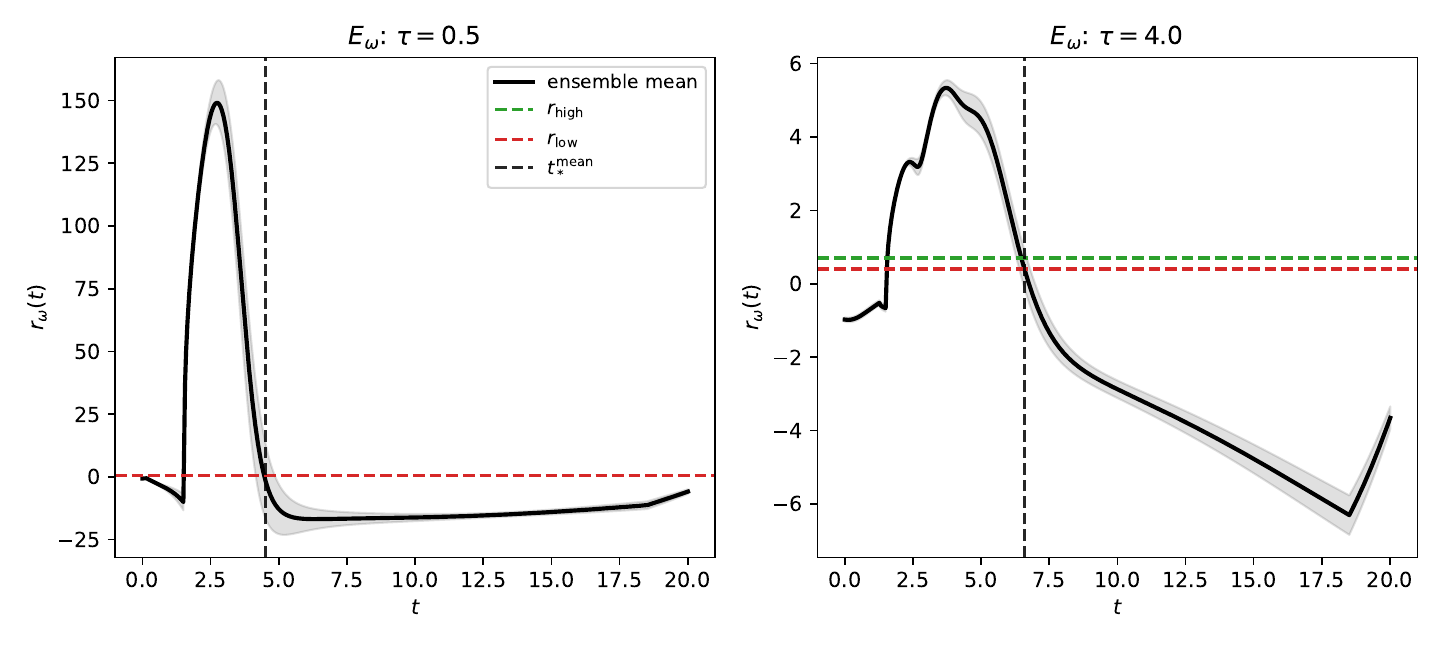}
    \caption{Frequency-based tracking ratio \(r_\omega(t)\) for representative fast and slow protocols. The dashed horizontal lines indicate the thresholds used in the operational extraction of \(t_*\), and the vertical line marks the ensemble-mean freeze-out time. The slow protocol delays the loss of coherent tracking substantially relative to the fast one.}
    \label{fig:ratiopanels}
\end{figure}

To understand the underlying dynamics, Fig.~\ref{fig:ratiopanels} shows the corresponding tracking ratio \(r_\omega(t)\) for representative fast and slow protocols. The fast-ramp case exhibits an early sharp transient followed by a rapid loss of tracking, whereas the slow-ramp case remains close to zero for much longer before departing.

\begin{figure}[t]
    \centering
    \includegraphics[width=0.72\textwidth]{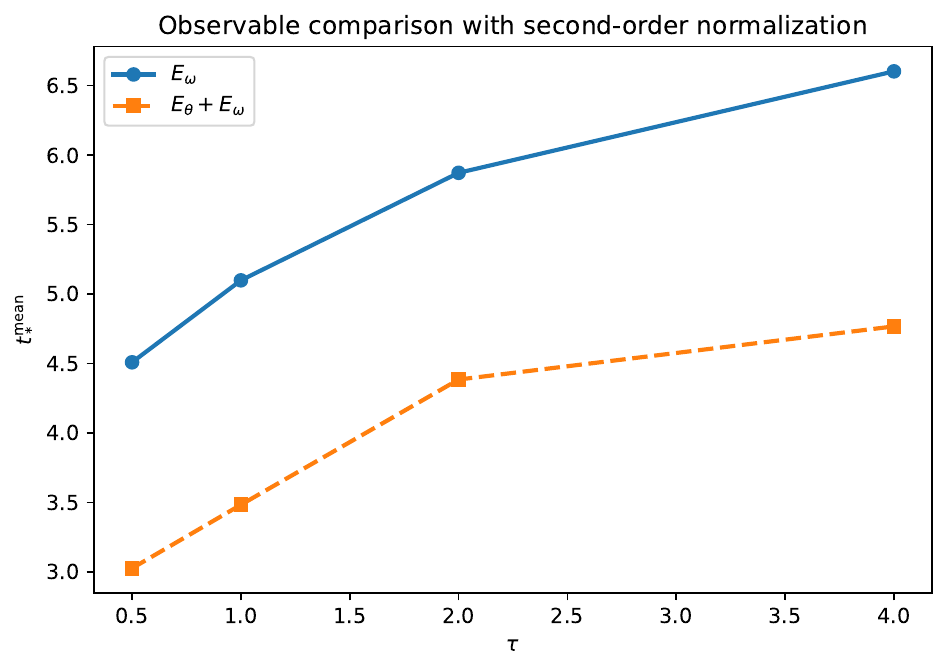}
    \caption{Comparison of operational freeze-out times obtained from different observables. The frequency-disagreement observable \(E_\omega\) produces the clearest and most monotonic protocol dependence. The combined observable \(E_\theta+E_\omega\) shows a similar but weaker trend, while the phase-sector observable by itself is less informative in the second-order regime.}
    \label{fig:observable_compare}
\end{figure}

Finally, Fig.~\ref{fig:observable_compare} compares the freeze-out time extracted from three candidate observables: the phase-disagreement energy, a combined phase-plus-frequency quantity, and the pure frequency-disagreement energy \(E_\omega\). The frequency sector gives the cleanest and most monotonic dependence on \(\tau\), while the phase sector remains weaker and more ambiguous. This motivates our choice of \(E_\omega\) as the primary observable for the rest of the paper.

In additional checks, the monotonic dependence of the extracted freeze-out time on \(\tau\) remained stable under moderate variations of the smoothing window and threshold choices entering the operational definition of \(t_*\). These changes shift the absolute calibration of the crossover time, but they do not alter the qualitative protocol dependence.

This is the first main positive result of the paper: once the observable and normalization are chosen in a way appropriate to the second-order problem, coherent-tracking loss becomes a robust and protocol-sensitive crossover phenomenon.

\section{Scaling with system size and adjacency structure}
\label{sec:results_scaling}

\subsection{Scaling with the number of nodes}

\begin{figure}[t]
    \centering
    \includegraphics[width=0.72\textwidth]{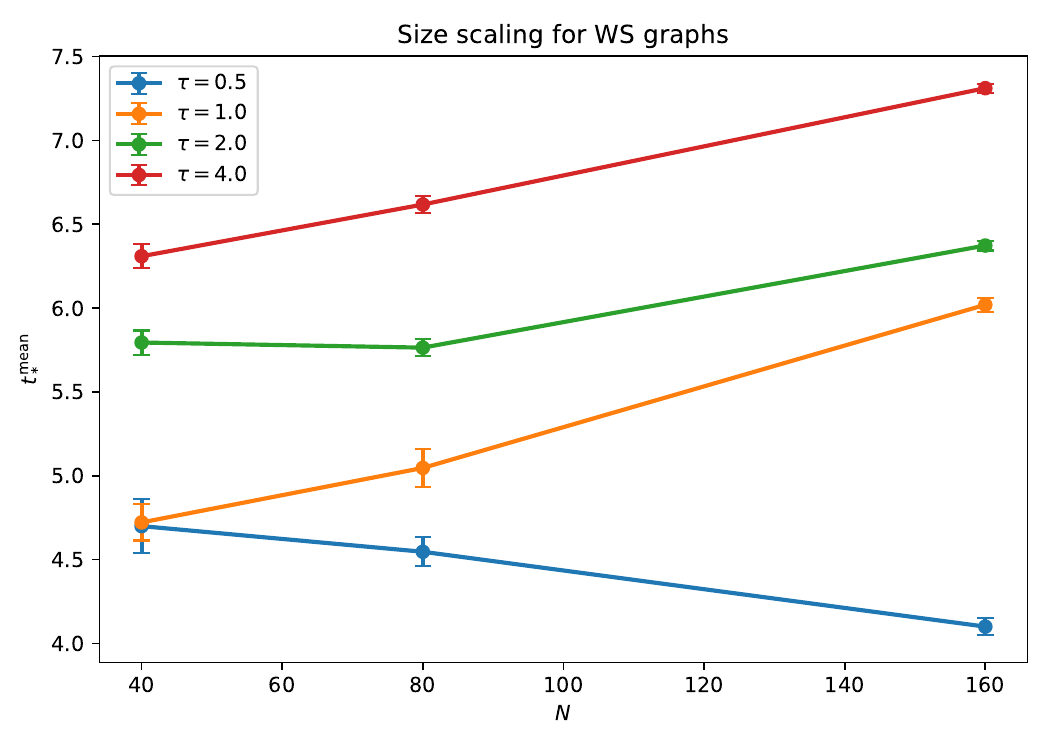}
    \caption{Size scaling of the frequency-based freeze-out time for WS graphs. The operational crossover remains well defined across system sizes and retains a clear dependence on the protocol timescale.}
    \label{fig:n_scaling_tstar}
\end{figure}

A central question is whether the frequency-based diagnostic is a finite-size artifact or a robust collective quantity. Figure~\ref{fig:n_scaling_tstar} shows the freeze-out time as a function of system size for WS graphs and several protocol timescales. The operational crossover remains well defined as \(N\) grows and retains its monotonic dependence on \(\tau\).

\begin{figure}[t]
    \centering
    \includegraphics[width=0.72\textwidth]{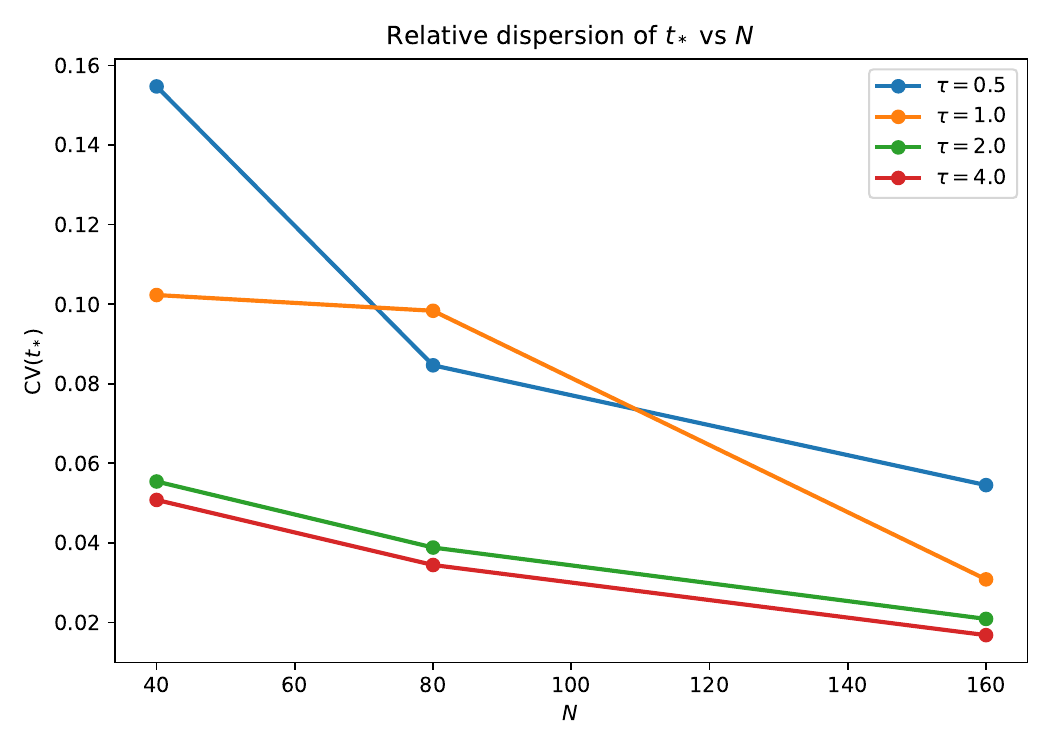}
    \caption{Coefficient of variation of the frequency-based freeze-out time as a function of system size for WS graphs. The decrease with \(N\) indicates that the diagnostic becomes sharper and less noisy in larger networks. In the fast-ramp case, the initial spike reflects a launch-transient effect of the logarithmic derivative and does not by itself define the operational freeze-out event.}
    \label{fig:n_scaling_cv}
\end{figure}

Even more importantly, the relative dispersion decreases with size. Figure~\ref{fig:n_scaling_cv} shows the coefficient of variation of \(t_*\), which drops systematically as \(N\) increases. Thus the \(E_\omega\)-based diagnostic becomes sharper and more self-averaging in larger networks.

\subsection{Dependence on adjacency structure}

\begin{figure}[t]
    \centering
    \includegraphics[width=0.78\textwidth]{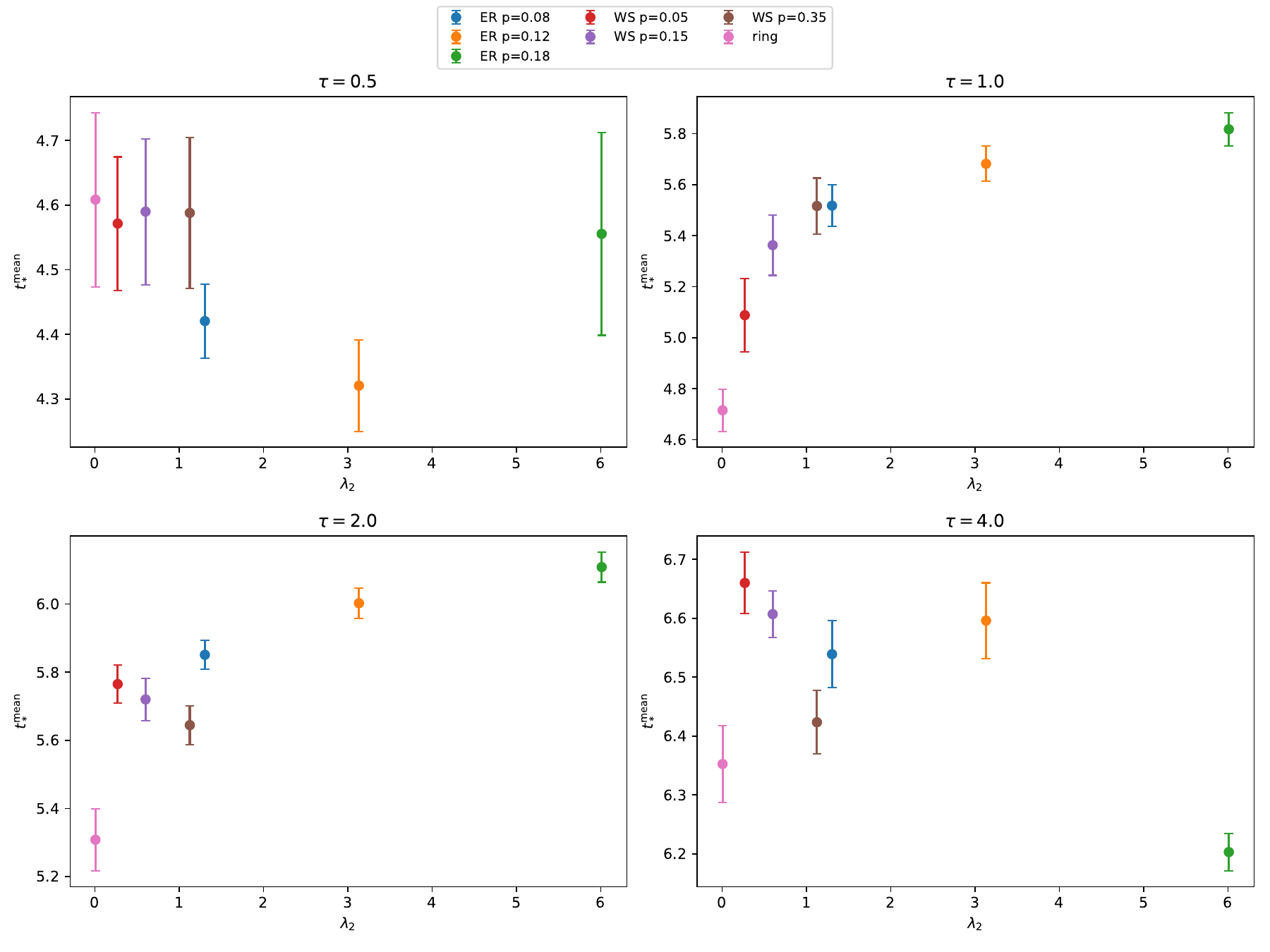}
    \caption{Frequency-based freeze-out time versus algebraic connectivity \(\lambda_2\) for ER, WS, and ring graphs at fixed \(N\), shown for several protocol timescales. The dependence is clearly structured but does not collapse to a universal one-parameter function of \(\lambda_2\).}
    \label{fig:tstar_lam2}
\end{figure}

We next examine how the operational freeze-out time depends on the structure of \(A_{ij}\). Figure~\ref{fig:tstar_lam2} plots \(t_*\) against the algebraic connectivity \(\lambda_2\) for ER, WS, and ring graphs at fixed \(N\). There is a clear spectral trend: graphs with larger \(\lambda_2\) tend to exhibit different tracking-loss times than graphs with very small \(\lambda_2\). However, the data do not collapse onto a single universal curve. Graph families with comparable average degree or comparable \(\lambda_2\) can still differ noticeably in \(t_*\).

Alternative scalar summaries such as mean degree lead to the same qualitative conclusion: simple graph-level descriptors organize the data only partially.

Taken together, these results show that the second-order frequency-sector freeze-out time is influenced by adjacency structure in a way that is only partly captured by \(\lambda_2\). The algebraic connectivity remains an important organizing descriptor, but topology beyond the slowest spectral scale matters.

\subsection{Beyond \(\lambda_2\): additional structural descriptors}

To sharpen the previous conclusion, we tested a small set of graph-level descriptors beyond \(\lambda_2\). In addition to the algebraic connectivity itself, we consider the low-spectrum spacing
\[
\Delta_{23}:=\lambda_3-\lambda_2,
\]
the ratio
\[
\rho_{23}:=\lambda_3/\lambda_2,
\]
the mean path length, the clustering coefficient, and the inverse participation ratio of the Fiedler eigenvector \(u^{(2)}\),
\[
\mathrm{IPR}_2:=\frac{\sum_i (u^{(2)}_i)^4}
{\left(\sum_i (u^{(2)}_i)^2\right)^2}.
\]
Larger \(\mathrm{IPR}_2\) corresponds to stronger localization of the slowest nontrivial Laplacian mode.

The motivation for including \(\mathrm{IPR}_2\) is simple. If the slowest mode is strongly localized, then the bottleneck controlling the network's slowest collective response is concentrated on a smaller subset of nodes, so the operational freeze-out time may depend not only on the value of the slowest spectral scale but also on how spatially distributed that mode is across the graph.

The goal here is not to build a predictive regression model, but to determine which simple descriptors materially improve the explanation of graph-to-graph variation in \(t_*\). The regressions reported below should therefore be read as exploratory diagnostics of structural relevance, not as final predictive models.

The results show a clear regime dependence. In the fast-ramp regime, none of the tested descriptors explains much of the variance, indicating that the crossover is dominated by transient effects rather than static graph structure. At intermediate ramps, by contrast, \(\lambda_2\) already has substantial explanatory power, but the inclusion of one extra structural descriptor improves the picture significantly. Among the candidates tested, the strongest single refinement is the Fiedler-mode localization measure \(\mathrm{IPR}_2\), while clustering and path length also help. In the slow-ramp regime, the dependence on \(\lambda_2\) weakens somewhat, but low-spectrum structure still matters, with \(\lambda_3/\lambda_2\) providing the cleanest single improvement beyond \(\lambda_2\).

This is an important refinement of the topology story. The graph dependence of \(t_*\) is not arbitrary, nor is it exhausted by one scalar descriptor. Instead, the operational freeze-out time is organized by a small set of structural quantities whose relevance depends on the protocol regime: transient-dominated fast ramps, localization-sensitive intermediate ramps, and slower-ramp regimes in which the shape of the low end of the spectrum becomes more relevant.

\begin{figure}[t]
    \centering
    \includegraphics[width=0.78\textwidth]{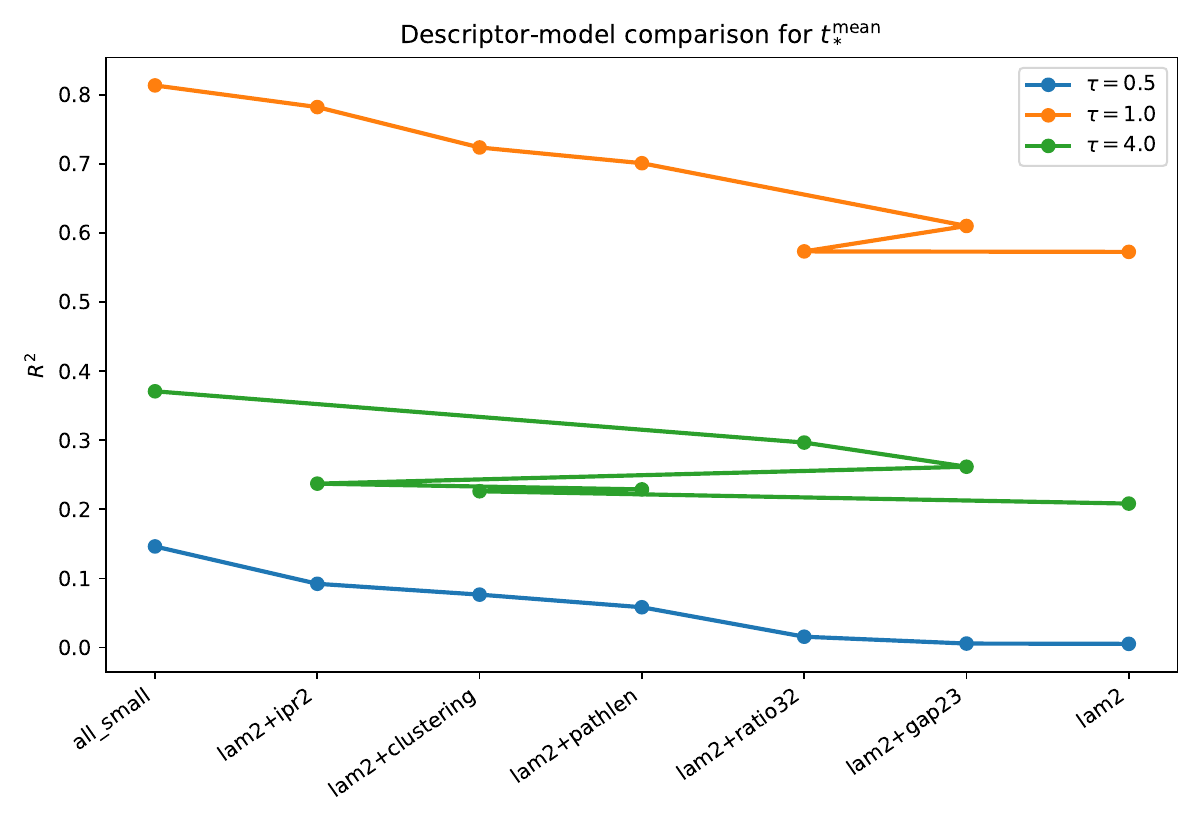}
    \caption{Comparison of simple descriptor models for the graph-to-graph variation of the frequency-based freeze-out time. At fast ramps, no static descriptor performs well. At intermediate ramps, \(\lambda_2\) already explains a substantial fraction of the variation, but the inclusion of additional structural information---especially Fiedler-mode localization---improves the description significantly. At slower ramps, low-spectrum structure such as \(\lambda_3/\lambda_2\) becomes the most useful single refinement beyond \(\lambda_2\).}
    \label{fig:model_comparison}
\end{figure}

\section{Slow-ramp envelopes and the limits of diagonal low-mode closures}
\label{sec:envelopes}

\begin{figure}[t]
    \centering
    \includegraphics[width=0.72\textwidth]{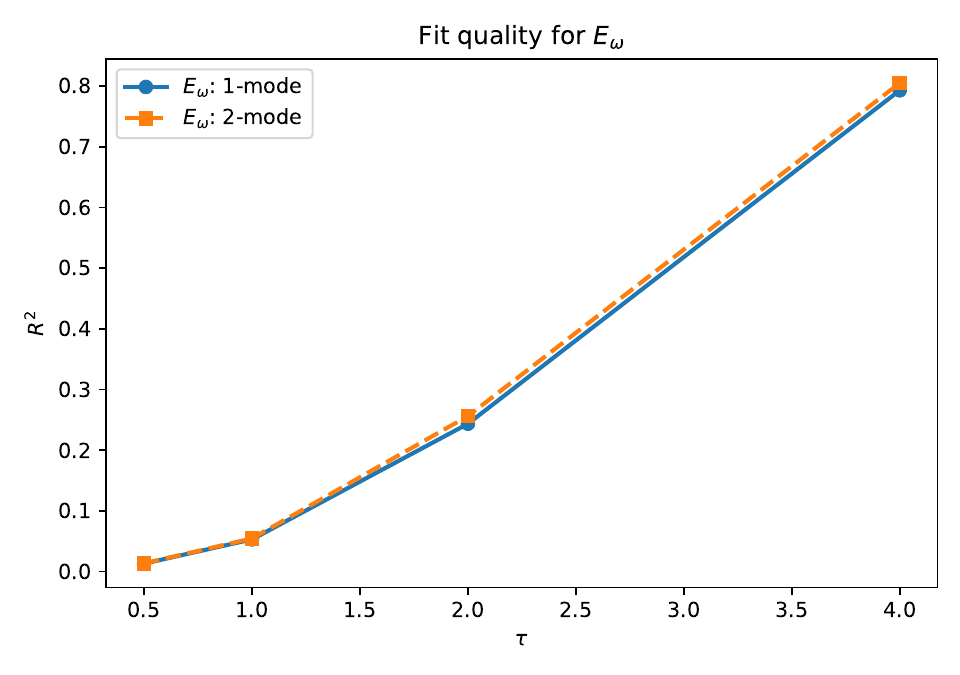}
    \caption{Fit quality for the one-mode and two-mode diagonal-envelope descriptions of \(E_\omega(t)\). The second-order envelope improves significantly in the slow-ramp regime, but remains weak for faster protocols.}
    \label{fig:eomega_quality}
\end{figure}

A natural secondary question is whether the same second-order modal rate that normalizes the tracking ratio can also support a quantitative relaxation theory. The associated fit-quality summary is shown in Fig.~\ref{fig:eomega_quality}.

The results are mixed. The second-order diagonal-envelope picture is not uniformly successful across the full protocol range: fit quality is poor in the fast-ramp regime and improves only progressively as the ramp slows. By \(\tau=4\), the description becomes reasonable, but at smaller \(\tau\) it is not a convincing quantitative closure. The improvement from one mode to two modes is real but modest.

This should not be viewed as a contradiction of the operational tracking result. Rather, it establishes a separation between two roles played by the second-order modal rate. As a \emph{local reference scale}, it is exactly what is needed to normalize the frequency-based tracking diagnostic. As a \emph{global diagonal relaxation envelope}, however, it is only asymptotically useful in the slower-ramp regime. In the faster regime, additional transient and nonlinear effects remain important. For this reason we regard the envelope analysis as a secondary, regime-dependent indication rather than as a coequal pillar of the paper's main contribution.

\section{Discussion}
\label{sec:discussion}

The main result of the paper is not a universal relaxation theory for all second-order observables. It is a more selective and, in our view, more useful statement. In this sense, the paper should be viewed as complementary to the classical synchronization and transient-stability literature on Kuramoto-type power-network reductions: rather than deriving new synchronization criteria, we isolate an operational diagnostic of \emph{tracking loss} under explicitly time-dependent driving.

In the second-order nonautonomous problem, the frequency sector provides the cleanest operational picture of coherent-tracking loss. A tracking ratio based on \(E_\omega\), normalized by the instantaneous second-order modal decay rate, yields a robust crossover time that depends systematically on the protocol, sharpens with system size, and is sensitive to graph structure. This makes it a promising reduced indicator of nonautonomous tracking loss in networked oscillator systems motivated by inverter-rich energy dynamics.

At the same time, the results expose the limits of the simplest reduced spectral picture. Graph topology matters beyond a single scalar descriptor such as \(\lambda_2\), and the most informative correction is not universal across protocols: localization-sensitive quantities are most useful at intermediate ramps, while low-spectrum spacing matters more in the slower-ramp regime. Diagonal low-mode relaxation envelopes are only asymptotically successful, and the phase sector exhibits strongly non-monotonic underdamped behavior that resists the naive low-mode closures inherited from first-order intuition.

This dual outcome should be seen as a gain in clarity rather than a weakness. It identifies the sector in which a reduced operational diagnostic is already reliable and the sector in which a stronger theory is still needed. In future work, the natural next steps are to clarify the role of the low end of the spectrum more systematically, to determine whether a minimal two-descriptor structural theory of \(t_*\) can be formulated, and to investigate whether richer modal-interaction corrections can improve the slow-ramp envelope picture.

From the point of view of applications, the present paper should be read as a bridge. It does not provide an engineering tool for realistic power-system operation. It does show, however, that a frequency-based reduced indicator of tracking loss can be defined in a nonautonomous second-order network model and that this indicator exhibits robust scaling with protocol, size, and topology. That is precisely the kind of reduced dynamical ingredient one would want before moving toward richer inverter-dominated network models.

\section{Conclusion}
\label{sec:conclusion}

We have introduced and tested a frequency-based operational diagnostic of coherent-tracking loss in nonautonomous second-order oscillator networks. The diagnostic is built from the frequency-disagreement observable \(E_\omega\) and normalized by the instantaneous second-order modal decay rate. In this form it yields a clear and robust protocol-dependent freeze-out time.

The resulting operational crossover sharpens with increasing system size, confirming that the signal is not a fragile finite-size artifact. It also depends systematically on graph structure. The algebraic connectivity \(\lambda_2\) provides an important first organizing descriptor, but the graph dependence is better understood through a small set of low-spectrum and structural quantities, especially Fiedler-mode localization at intermediate ramps and low-spectrum spacing in the slower-ramp regime. Spatially constrained graphs such as rings form a distinct non-generic regime.

Finally, we have shown that the same second-order modal scale that works well as a local normalization principle does not yet provide a universally accurate diagonal low-mode relaxation closure. The central lesson is therefore sector-dependent: in the second-order nonautonomous problem the frequency sector supports a robust reduced operational description of tracking loss, whereas the phase sector does not. The latter remains much less reducible and appears to require additional structural information and richer, likely non-diagonal, dynamics beyond the simplest low-mode picture.

Beyond the specific oscillator model studied here, the present results suggest a broader class of application domains in which reduced diagnostics of nonautonomous tracking loss may be useful. The most immediate examples are converter-rich power systems and microgrids, where reduced inertia, reconfiguration, and time-dependent operating conditions make frequency tracking more relevant than static synchronization alone. More generally, the same logic may apply to networked control architectures in which couplings or effective gains evolve in time, including coordinated inverter arrays, electromechanical oscillator networks, and possibly multi-agent systems such as robotic or drone swarms operating under time-varying communication or control conditions. In all such cases, the value of the present study is not that it provides an engineering-ready prediction tool, but that it identifies a compact dynamical diagnostic principle: in second-order nonautonomous networks, loss of coherent tracking may be detected most cleanly in the frequency sector, and its dependence on network structure is only partly reducible to a single spectral scale.
\bibliographystyle{apsrev4-2}
\bibliography{refs}
\end{document}